\documentclass{article}
\usepackage{siunitx} 
\usepackage[english]{babel}
\usepackage{amsmath}  
\usepackage{amssymb}  
\usepackage{multicol}
\usepackage{float}
\usepackage{siunitx} 
\usepackage{booktabs} 
\usepackage{braket}
\usepackage{cite}
\usepackage{fancyhdr}

\newcommand{\sg}{\operatorname{sg}} 

\usepackage[letterpaper,top=1.7cm,bottom=2cm,left=1.5cm,right=1.5cm,marginparwidth=1.75cm]{geometry}

\usepackage{amsmath}
\usepackage{graphicx}
\usepackage[colorlinks=true, allcolors=blue]{hyperref}
\usepackage{fancyhdr} 

\fancypagestyle{myfooter}{%
  \fancyhf{} 
  
  \fancyfoot[L]{Email: \textit{ycy1387@163.com}} 
}

\title{Latent Space Single-Pixel Imaging Under Low-Sampling Conditions}

\author{Chenyu Yuan \\
[3mm]\small\sl Department of Physics, University of Shanghai for Science and Technology, Shanghai, 200093, China}

\begin{document}
\maketitle
\thispagestyle{myfooter} 
\begin{abstract}
In recent years, the introduction of deep learning into the field of single-pixel imaging has garnered significant attention. However, traditional networks often operate within the pixel space. To address this, we innovatively migrate single-pixel imaging to the latent space, naming this framework LSSPI (Latent Space Single-Pixel Imaging). Within the latent space, we conduct in-depth explorations into both reconstruction and generation tasks for single-pixel imaging. Notably, this approach significantly enhances imaging capabilities even under low sampling rate conditions. Compared to conventional deep learning networks, LSSPI not only reconstructs images with higher signal-to-noise ratios (SNR) and richer details under equivalent sampling rates but also enables blind denoising and effective recovery of high-frequency information. Furthermore, by migrating single-pixel imaging to the latent space, LSSPI achieves superior advantages in terms of model parameter efficiency and reconstruction speed. Its excellent computational efficiency further positions it as an ideal solution for low-sampling single-pixel imaging applications, effectively driving the practical implementation of single-pixel imaging technology.
\end{abstract}

\begin{multicols}{2}
\section{Introduction}
Single-pixel imaging (SPI) \cite{sunCollectiveNoiseModel2018,wanDemonstrationAsynchronousComputational2022,gongPerformanceComparisonComputational2022,vasileSinglePixelSensing2016,heEnergyselectiveXrayGhost2020}, an emerging computational imaging modality, reconstructs images from bucket detector signals by leveraging the second-order correlation properties of quantum or classical light. This technique collects photons interacting with the object and demonstrates notable advantages in detection sensitivity, dark count suppression, and spectral range extension. Over the past decade, these strengths have driven the continuous growth of SPI applications in fields such as remote sensing \cite{gongThreedimensionalGhostImaging2016,wangAirborneInfraredThreedimensional2018}, 3D imaging \cite{sun3DComputationalImaging2013,sunSinglepixelThreedimensionalImaging2016}, terahertz imaging \cite{maTerahertzSinglePixel2012,shrekenhamerTerahertzSinglePixel2013,stantchevRealtimeTerahertzImaging2020}, and optical encryption   \cite{yuJointAuthenticationPublic2024}. Nevertheless, SPI inherently requires a large number of measurements to reconstruct high-resolution images. The trade-off between acquisition time and image quality has constrained its broader application prospects.

To overcome this limitation, the academic community has been dedicated to exploring optimization algorithms for reducing sampling rates \cite{katzCompressiveGhostImaging2009,lyuDeeplearningbasedGhostImaging2017}. The emergence of Compressed Sensing (CS) theory has led to significant breakthroughs in this field, effectively enabling high-quality image reconstruction under low sampling rates \cite{duarteSinglepixelImagingCompressive2008}. However, CS technology relies on image sparsity and iterative optimization, suffering from high computational complexity—particularly pronounced under ultra-low sampling conditions \cite{qiuComprehensiveComparisonSinglepixel2020}. In recent years, data-driven deep learning (DL) methods have been introduced into the SPI field, significantly improving the quality of reconstructed images. Current DL-based SPI reconstruction methods are primarily categorized into two types: deterministic reconstruction models and probabilistic reconstruction models.

Deterministic reconstruction models \cite{wangLearningSimulationEndtoend2019,heGhostImagingBased2018,wangSinglepixelImagingUsing2021,wangFarfieldSuperresolutionGhost2022} learn the complex mapping from bucket detector signals to images through neural networks, offering simplicity in training and high operational efficiency. However, they often suffer from loss of high-frequency information and are prone to noise issues. Probabilistic reconstruction models \cite{zhaoHighqualityComputationalGhost2023,maoHighqualityHighdiversityConditionally2023} generate realistic images guided by bucket detector signals but exhibit limited output controllability: traditional Generative Adversarial Networks (GANs) \cite{goodfellowGenerativeAdversarialNets2014} suffer from defects such as mode collapse, leading to suboptimal image quality. Although Denoising Diffusion Probabilistic Models (DDPMs) \cite{hoDenoisingDiffusionProbabilistic2020} were proposed to address some training challenges of GANs, their image reconstruction process requires multiple iterative sampling steps, resulting in limited reconstruction efficiency that struggles to meet real-time application requirements. Moreover, all the above methods perform image reconstruction in the pixel space, where model parameters increase significantly with higher image resolutions, thereby causing a substantial surge in computational resource demands.

In this article, we migrate the single-pixel imaging reconstruction process to the latent space, effectively reducing the model training burden and shortening the image reconstruction time. Within the latent space, we integrate the strengths of both deterministic and probabilistic reconstruction models, achieving effective recovery of high-frequency information and enabling blind denoising functionality. This significantly enhances the visual quality of reconstructed images. Furthermore, we explore the application of bucket detector signals in image generation, discovering that they possess a guidance capability analogous to natural language. This capability allows them to guide the generation of images with specific features, thereby expanding the application scope of single-pixel imaging technology.

\end{multicols}

\begin{multicols}{2}
\section{Method}
\begin{figure}[htbp]
  \centering
  \includegraphics[width=0.7\textwidth]{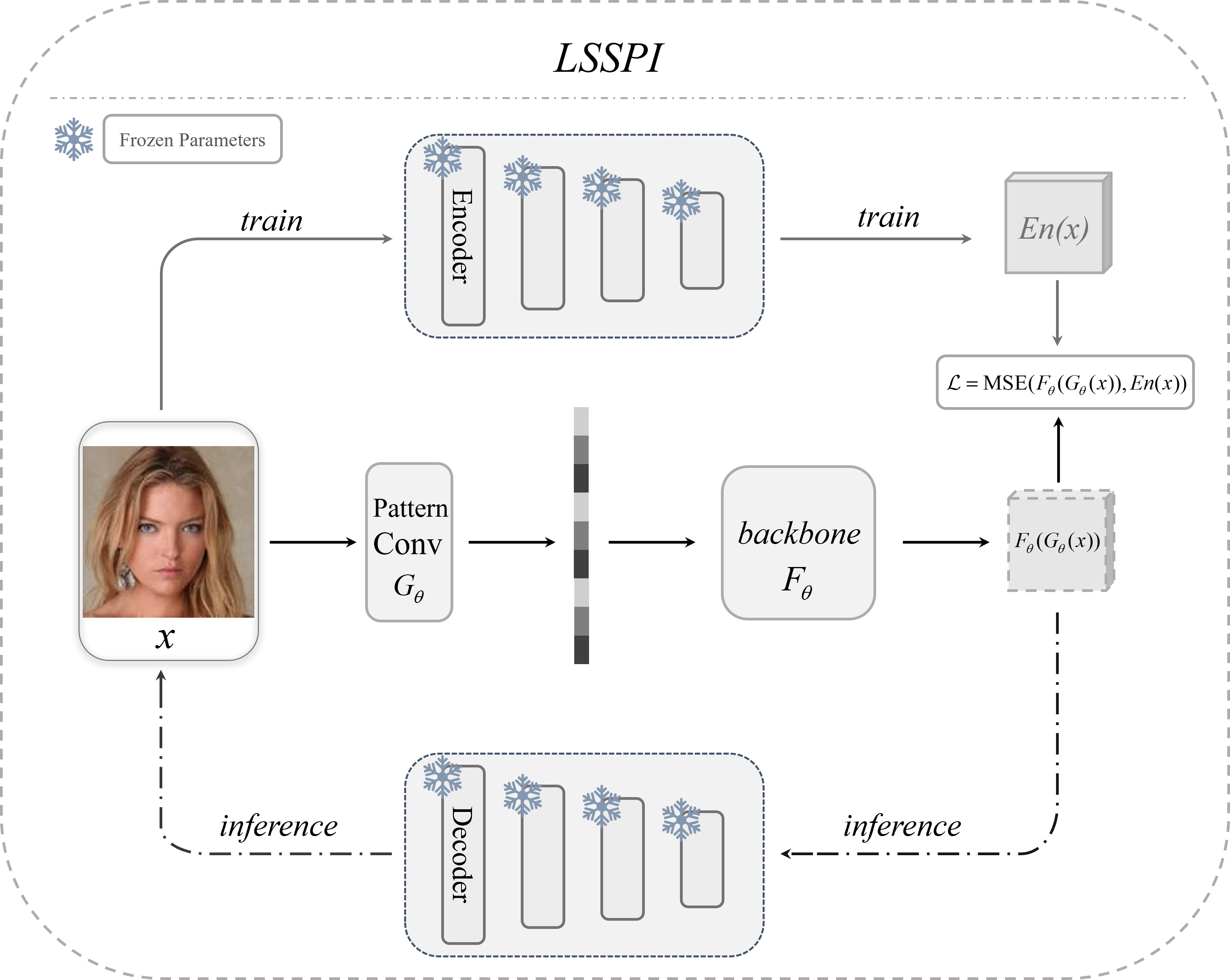} 
  \caption{Schematic Diagram of LSSPI.} 
  \label{fig:one} 
\end{figure}

The main framework of the proposed method is illustrated in Fig. \ref{fig:one}. This approach is a self-supervised training method that requires no additional labels; it only needs a pre-trained Variational Autoencoder (VAE) \cite{kingmaAutoencodingVariationalBayes2013,rombachHighResolutionImageSynthesis2022a} to compress images into latent space vectors. In conventional single-pixel deep learning methods, speckle patterns are typically used to encode images, while deep learning networks are employed to decode bucket detector signals. In our method, however, the deep learning network serves as an encoder to re-encode the bucket detector signals, and the decoding process leverages the decoder of the pre-trained VAE. This thereby migrates single-pixel imaging into the latent space. An obvious advantage of this approach is its ability to effectively reduce the size of deep learning model parameters. By compressing images into latent space vectors (e.g., compressing 64×64 resolution into 16×16×8), the output results and model parameters of the deep learning network are correspondingly reduced in scale.
\end{multicols}

\begin{multicols}{2}
\subsection{ViT}
\begin{figure}[htbp]
  \centering
  \includegraphics[width=0.7\textwidth]{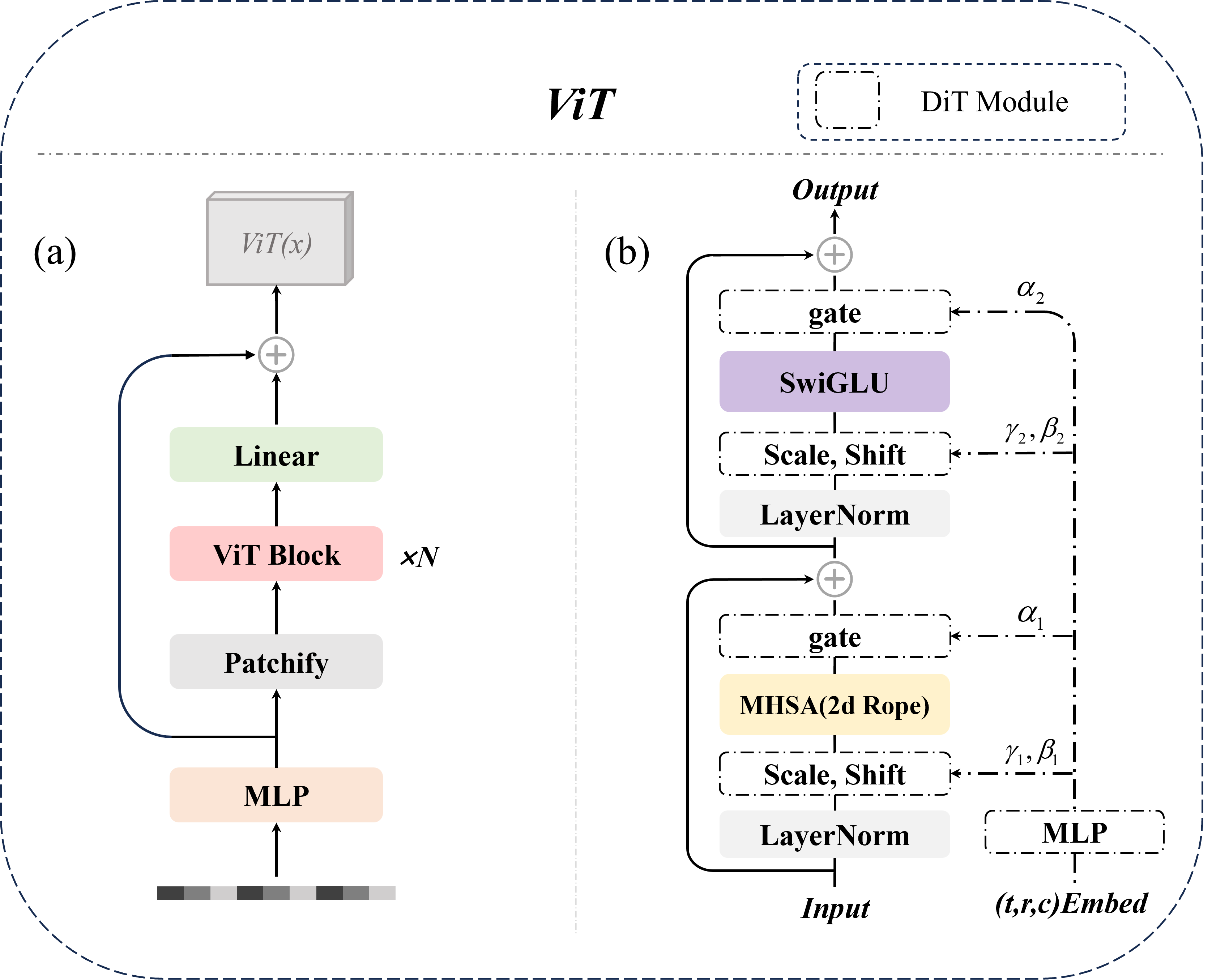} 
  \caption{Reconstruction Model Based on ViT. (a) ViT-based network architecture. (b) ViT/DiT block modules.} 
  \label{fig:two} 
\end{figure}

The architecture of the deep learning network used to establish the mapping from bucket detector signals to the target is illustrated in Fig. \ref{fig:two}. The overall network structure is highly concise: the Multilayer Perceptron (MLP) is responsible for encoding the dimensionality of bucket detector signals into the latent space vector dimensionality. For feature extraction, we adopt the Vision Transformer (ViT) \cite{dosovitskiyImageWorth16x162021} architecture, primarily due to its prominent global receptive field and scalability advantages. The global receptive field ensures that the model can fully comprehend the overall structure of the image as well as the semantic correlations between distant elements. The excellent scalability of ViT further allows us to flexibly select an appropriate model scale based on the computational resource constraints and accuracy requirements of the task. Additionally, it facilitates our use of larger-scale pre-trained models for transfer learning, thereby achieving the optimal balance between efficiency and performance.
\end{multicols}
\begin{multicols}{2}
\subsection{Flow Matching}

Flow Matching (FM) \cite{lipmanFlowMatchingGenerative2022,liuFlowStraightFast2022a} is a class of generative models that learn to match the flow represented by the velocity field between two probability distributions. Formally, given data $x \sim p_{\text{data}}(x)$ and a prior distribution $\epsilon \sim p_{\text{prior}}(\epsilon)$, a flow path can be constructed as $x_t = \alpha_t x + b_t \epsilon$, where $t$ is the time variable. A common approach is to take $\alpha_t = 1 - t$ and $b_t = t$, then the velocity is naturally defined as Eq.~\ref{eq1}
\begin{equation}
\frac{\mathrm{d}x_{t}}{\mathrm{d}t} = \epsilon - x
\label{eq1}
\end{equation}

However, the above expression is non-causal. We therefore require a model to predict velocity using the current state. Consequently, the loss function is given by Eq.~\ref{eq2}.

\begin{equation}
\mathcal{L} = \int_{0}^{1} \mathbb{E}_{x,\epsilon} \left[ \left\| \epsilon - x - v_{\theta} \left( x_{t}, t \right) \right\|^{2} \right] dt
\label{eq2}
\end{equation}

After training is completed, given the prior $\epsilon$, $x$ can be obtained by solving the ordinary differential equation (ODE) in Eq.~\ref{eq3}
\begin{equation}
\frac{dx_{t}}{dt}=v_{\theta}(x_{t},t)
\label{eq3}
\end{equation}

However, solving the ODE may not yield desirable results since it represents an unconditional generative model. To effectively control the generated outcomes, the natural approach involves incorporating conditions into the velocity model.

\begin{equation}
    \begin{gathered}
    x_{c} = \text{ViT}(c)  \\
    \mathcal{L} = \int_{0}^{1} \mathbb{E}_{x, \epsilon} \left[ \left\lVert \epsilon - x - v_{\theta}(x_t, t, x_c) \right\rVert^{2} \right] dt \\
    \frac{d x_{t}}{dt} = v_{\theta}(x_t, t, x_c) 
    \end{gathered}
\label{eq4}
\end{equation}

In Eq.~\ref{eq4}, $ c $ represents the bucket signal and is incorporated as a conditional input into the velocity model. After training is completed, the state $ x $ is obtained by solving the ODE given in Eq.~\ref{eq4}. Since this $ x $ is generated under the guidance of the condition $ x_c $, it will be closely correlated with $ x_c $.

\subsection{MeanFlow}
While MeanFlow (MF) \cite{gengMeanFlowsOnestep2025} demonstrates promising performance by establishing a conditioned velocity model and solving the corresponding ODE, the multi-step iterative process required for ODE solution incurs significant computational overhead. In our experiments, reconstructing a single image requires 20 seconds when the Number of Function Evaluations (NFE) is set to 160. This latency renders the approach prohibitively slow for real-time imaging applications. Consequently, there is a compelling need to develop conditioned velocity models capable of generating samples with drastically fewer steps, potentially even enabling single-step sampling. To address this limitation, we adopted the Mean Flow model. MF establishes an averaged velocity field, enabling efficient sample generation in multiple or even single steps.

Based on the physical definition of average velocity, we have Eq.~\ref{eq5}
\begin{equation}
    (t-r)u\big(x_{t},r,t\big)=\int_{r}^{t}\!\! v\big(x_{\tau},\tau\big)\tau
\label{eq5}
\end{equation}
$ u(x_t, r, t) $ represents the average velocity between time $ t $ and time $ r $, while $ v(x_{\tau}, \tau) $ denotes the instantaneous velocity. Taking the derivative with respect to time $ t $ on both sides simultaneously yields Eq.~\ref{eq6}.
\begin{equation}
    u\left( x_{t},r,t\right)=v\left( x_{t},t\right)-(t-r)\frac{d}{d t}u\left( x_{t},r,t\right)
\label{eq6}
\end{equation}
Thus, the training objective of Eq.~\ref{eq7} can be naturally derived.
\begin{equation}
    \begin{gathered}
        \frac{d}{dt}u\left( x_{t},r,t\right) = v\left( x_{t},t\right) \frac{\partial u}{\partial x_{t}} + \frac{\partial u}{\partial t} \\
        u_{\mathrm{target}} = v\left( x_{t},t\right) - (t-r)\left( v\left( x_{t},t\right) \frac{\partial u_{\theta}}{\partial x_{t}} + \frac{\partial u_{\theta}}{\partial t} \right) \\
        \mathcal{L}(\theta) = \mathbb{E} \left\| u_{\theta}\left( x_{t},r,t\right) - \sg\left( u_{\mathrm{target}} \right) \right\|_{2}^{2}
    \end{gathered}
\label{eq7}
\end{equation}
The sg() operator in Eq.~\ref{eq7} implements gradient stopping, serving dual purposes: preventing secondary backpropagation to reduce training computation, and avoiding label leakage in the learning process.

Therefore, this formulation similarly achieves unconditional generation. After incorporating the bucket signal condition $ c $, the training objective given in Eq.~\ref{eq8} can be expressed as:
\begin{equation}
    \begin{gathered}
        x_{c} = \mathrm{ViT}(c) \\
        \tilde{v}_{t} \triangleq \omega(\epsilon-x) + \kappa u_{\theta}^{\mathrm{cfg}}(x_{t}, t, t \mid c, x_{c}) + \\(1-\omega-\kappa) u_{\theta}^{\mathrm{cfg}}(x_{t}, t, t) \\
        u_{\mathrm{target}} = \tilde{v}_{t} - (t-r)\left( \tilde{v}_{t} \frac{\partial u_{\theta}^{\mathrm{cfg}}}{\partial x_{t}} + \frac{\partial u_{\theta}^{\mathrm{cfg}}}{\partial t} \right) \\
        \mathcal{L}(\theta) = \mathbb{E} \left\| u_{\theta}^{\mathrm{cfg}}(x_{t}, r, t \mid c, x_{c}) - \sg(u_{\mathrm{target}}) \right\|_{2}^{2}
    \end{gathered}
\label{eq8}
\end{equation}

Here, the Classifier-Free Guidance (CFG) technique \cite{hoClassifierfreeDiffusionGuidance2022} is employed as described in Eq.~\ref{eq8}, where $c$ denotes the bucket signal, and $w$ and $k$ represent guidance coefficients. Upon completion of training, both few-step sampling and single-step sampling (at $t=1, r=0$) can be achieved via Eq.~\ref{eq9}.

\begin{equation}
    x_{r}=x_{t}-(t-r)u_{\theta}(x_{t},r,t,c,x_{c})
\label{eq9}
\end{equation}

\subsection{ControlNet}

\begin{figure}[H]
  \centering
  \includegraphics[width=0.9\linewidth]{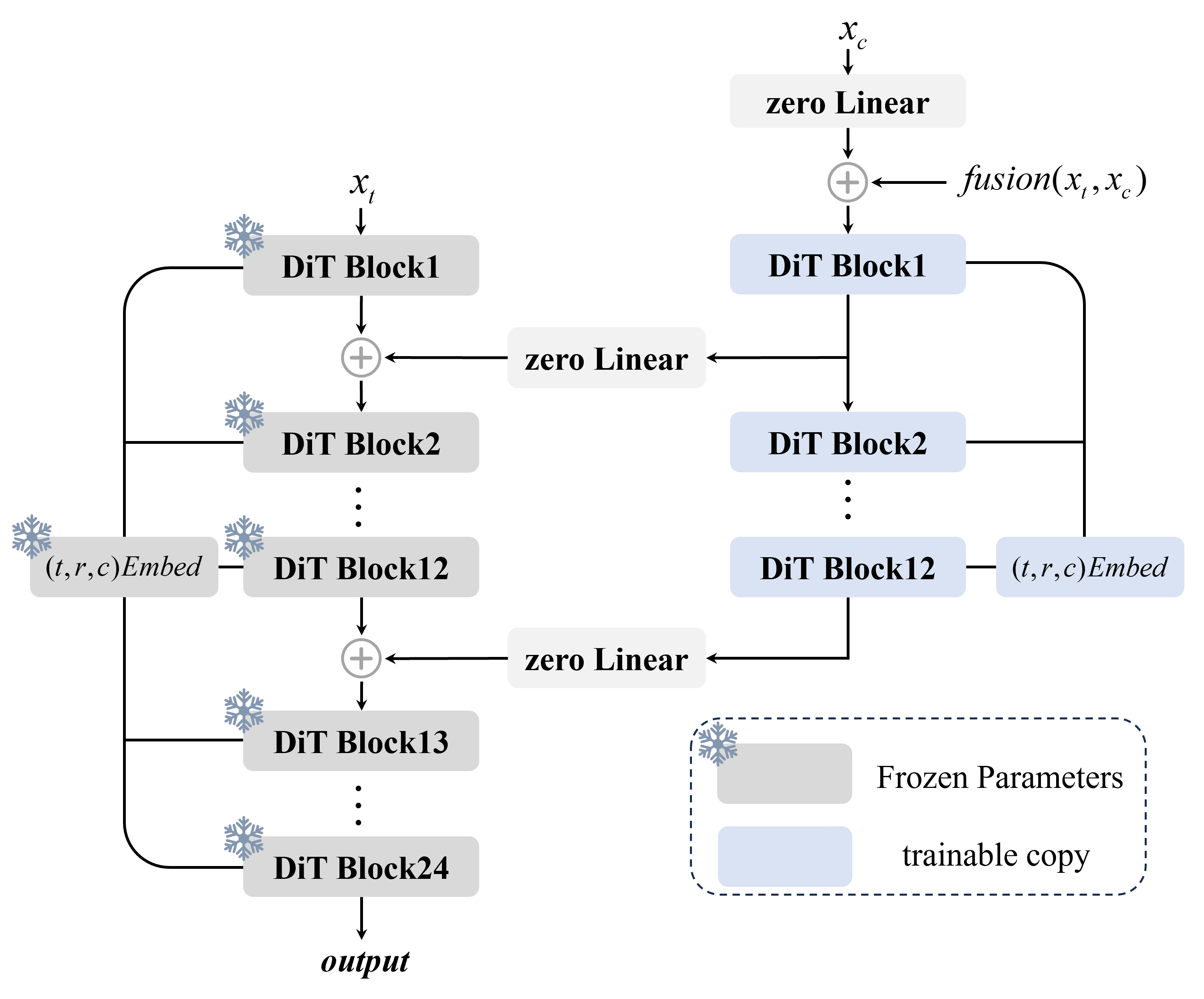}
  \caption{ControlNet network structure.}
  \label{fig:three}
\end{figure}

In both the Flow Matching and MeanFlow frameworks, a conditional variable $x_c$ is integrated. This conditional variable is injected through the ControlNet \cite{zhangAddingConditionalControl2023a} module, specifically achieved by further training on a pre-trained model (i.e., the portion with frozen parameters as shown in Fig. \ref{fig:three}). The backbone adopts a DiT \cite{peeblesScalableDiffusionModels2023} architecture (Fig. \ref{fig:one}(b)), and the overall model architecture is illustrated in Fig. \ref{fig:three}.  

Taking MeanFlow as an example, we first trained a conditional model $u_\theta(x_t, t, r, c)$. Subsequently, with the parameters of this initial model held fixed, we proceeded to train an extended conditional model $u_\theta(x_t, t, r, c, x_c)$.

The fusion operation and conditional encoding in the figure are formally defined as follows::

\begin{equation}
\text{fusion}(x_{t}, x_{c}) = \sigma \cdot \text{MLP}\big[ x_{t},\  \mu x_{c} \big]
\end{equation}
\begin{equation}
(t,r,c)Embed:=Embed(t)+Embed(t-r)+Embed(c)
\end{equation}
where $\sigma$ and $\mu$ are learnable constants, with $\sigma$ initialized to $1$ and $\mu$ initialized to $1 \times 10^{-4}$.
\subsection{Diversity generative}
Here, we briefly elaborate on the specific details of the diversity generation results, primarily focusing on the performance of the function $u_\theta(x_t, t, r, c)$ when $x_c$ is not used and the model is solely driven by the bucket signal $c$. The current experiments involve two training approaches; despite differences in their specific implementations, both methods achieve favorable generation performance. Their core distinction lies in whether the pretrained model is utilized during the encoding process of the bucket signal $c$.  

Taking CLIP \cite{radfordLearningTransferableVisual2021}, a widely used pretrained model, as an example, it is a multimodal contrastive learning model. Through pretraining on large-scale data, CLIP can establish associations between the bucket signal $c$ and latent space images. This associative capability enables CLIP to directly translate the bucket signal $c$ into a language understandable to the model, allowing it to serve as input features for direct participation in generation tasks or to accommodate the requirements of other downstream tasks without necessitating additional training. Leveraging this property, we next present the specific experimental results of the aforementioned two training approaches in diversity generation tasks.

\section{Results}
\subsection{Latent Space Parameters}
Since all subsequent models are trained within the latent space, we hereby present the relevant parameters of the latent space. All training processes were conducted on hardware consisting of an Intel Core i9-10980XE CPU, 32GB RAM, and an NVIDIA RTX 4090 GPU, with the PyTorch deep learning framework employed. The specific parameters used are provided in Table \ref{tab1} below.

\begin{table*}[htbp]
\centering
\caption{Latent Space Parameters}
\label{tab1}
\begin{tabular*}{0.9\textwidth}{@{\extracolsep{\fill}} c c c c @{}}
\toprule
\textbf{Pixel Dimensions} & \textbf{Latent Space Dimensions} & \textbf{Model Parameters} & \textbf{Compression Ratio} \\
\midrule
$64\times64$ & $16\times16\times8$ & 0.79M & 2 \\
$128\times128$ & $32\times32\times4$ & 0.77M & 4 \\
\bottomrule
\end{tabular*}
\end{table*}
\subsection{ViT Results}
\begin{figure*}[htbp]
  \centering
  \includegraphics[width=0.8\textwidth]{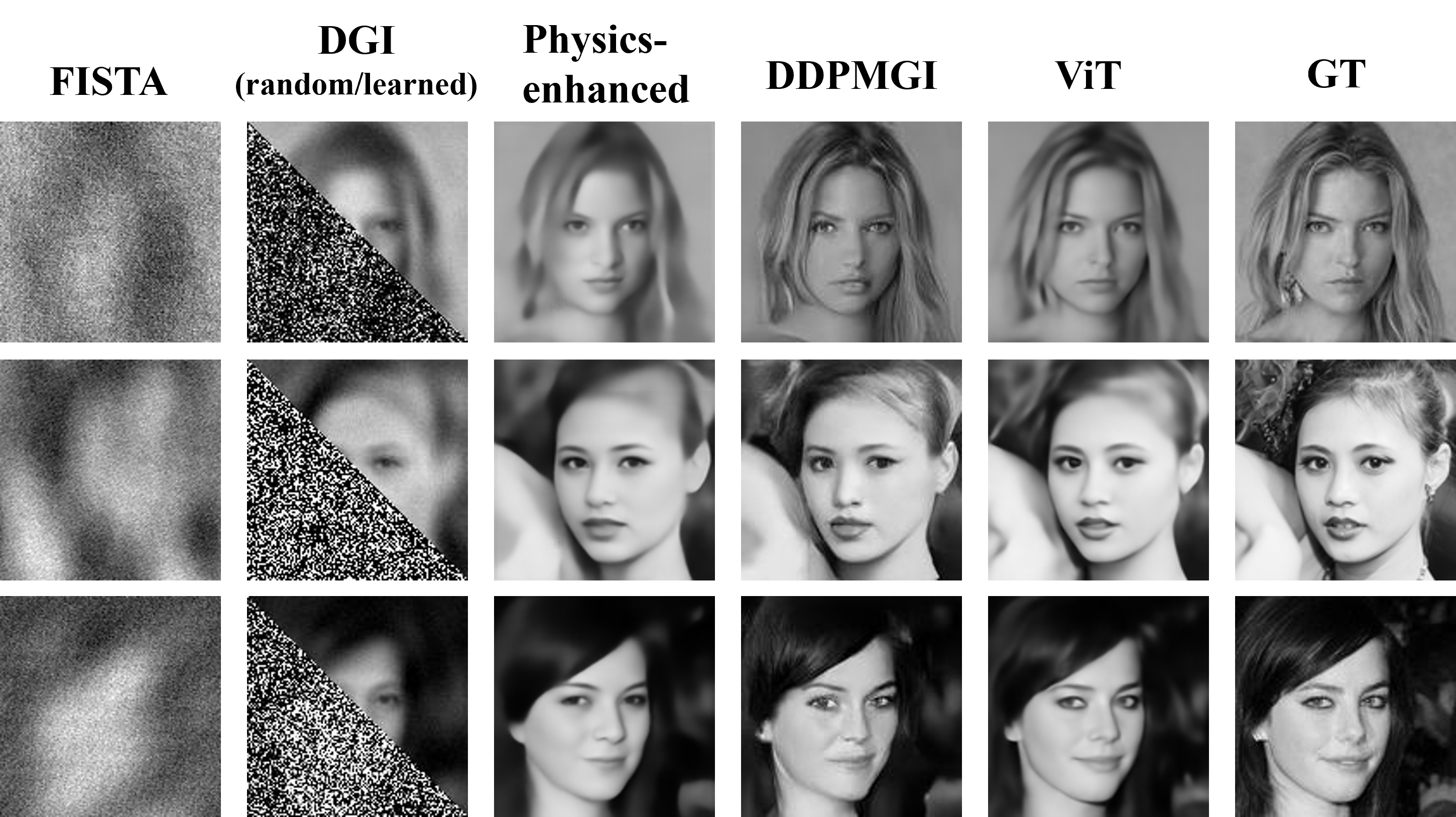} 
  \caption{Reconstruction results at a 4.8\% sampling rate.} 
  \label{fig:four} 
\end{figure*}

Here, we select representative single-pixel imaging models for comparison, namely DGI \cite{ferriDifferentialGhostImaging2010,wangSinglepixelImagingUsing2021}, FISTA \cite{beckFastIterativeShrinkagethresholding2009}, Physics-enhanced \cite{wangSinglepixelImagingUsing2021}, and DDPMGI \cite{maoHighqualityHighdiversityConditionally2023}. These methods represent compressive sensing, deterministic reconstruction, and probabilistic reconstruction approaches, respectively, and all perform reconstruction in the pixel domain. The training dataset employed is the Flickr-Faces-HQ Dataset. The corresponding results are presented in Fig. \ref{fig:four}.
\end{multicols}

\begin{multicols}{2}
We evaluated the aforementioned reconstruction methods on a dataset of 2000 images and computed the average Peak Signal-to-Noise Ratio (PSNR) and Structural Similarity Index (SSIM), as summarized in Table \ref{tab2}. The results indicate that the ViT network achieved superior performance in both PSNR and SSIM. Furthermore, owing to its reconstruction process being conducted in the latent space, the ViT model also demonstrated a significant advantage in reconstruction time compared to most models operating in the pixel domain.
\end{multicols}
\begin{table*}[htbp]
\centering
\caption{Comparison of Image Reconstruction}
\label{tab2}
\begin{tabular*}{0.9\textwidth}{@{\extracolsep{\fill}} c c c c @{}}
\toprule
\textbf{Model} & \textbf{PSNR} & \textbf{SSIM} & \textbf{Time} \\
\midrule
FISTA & 13.506 & 0.158 & 16s (iterations=50000) \\
DGI & 18.897 & 0.519 & 0.265s \\
Physics enhance & 21.769 & 0.722 & 0.406s \\
DDPMGI & 20.663 & 0.700 & 55s (NFE=500) \\
ViT & 23.668 & 0.811 & 0.312s \\
ViT (gray pattern) & 27.451 & 0.852 & 0.312s \\
\bottomrule
\end{tabular*}
\end{table*}
\begin{multicols}{2}
\subsection{Flow Matching and MeanFlow}
Although ViT-based reconstruction models have achieved favorable scores on the quantitative metrics PSNR and SSIM, it should be noted that their training relies on low-sampling data and high-compression-ratio measurement modes, and they are optimized using MSE (mean squared error) as the primary loss function. This combination objectively induces the models to tend toward generating pixel-average-optimal solutions, with side effects resulting in reconstructed images typically exhibiting over-smoothed apparent features, significant loss of visually critical high-frequency details (e.g., sharp edges and fine textures), and inevitable introduction of artifacts (e.g., blur artifacts, aliasing effects) and unstructured noise. 

\begin{figure}[htbp]
  \centering
  \includegraphics[width=0.8\textwidth]{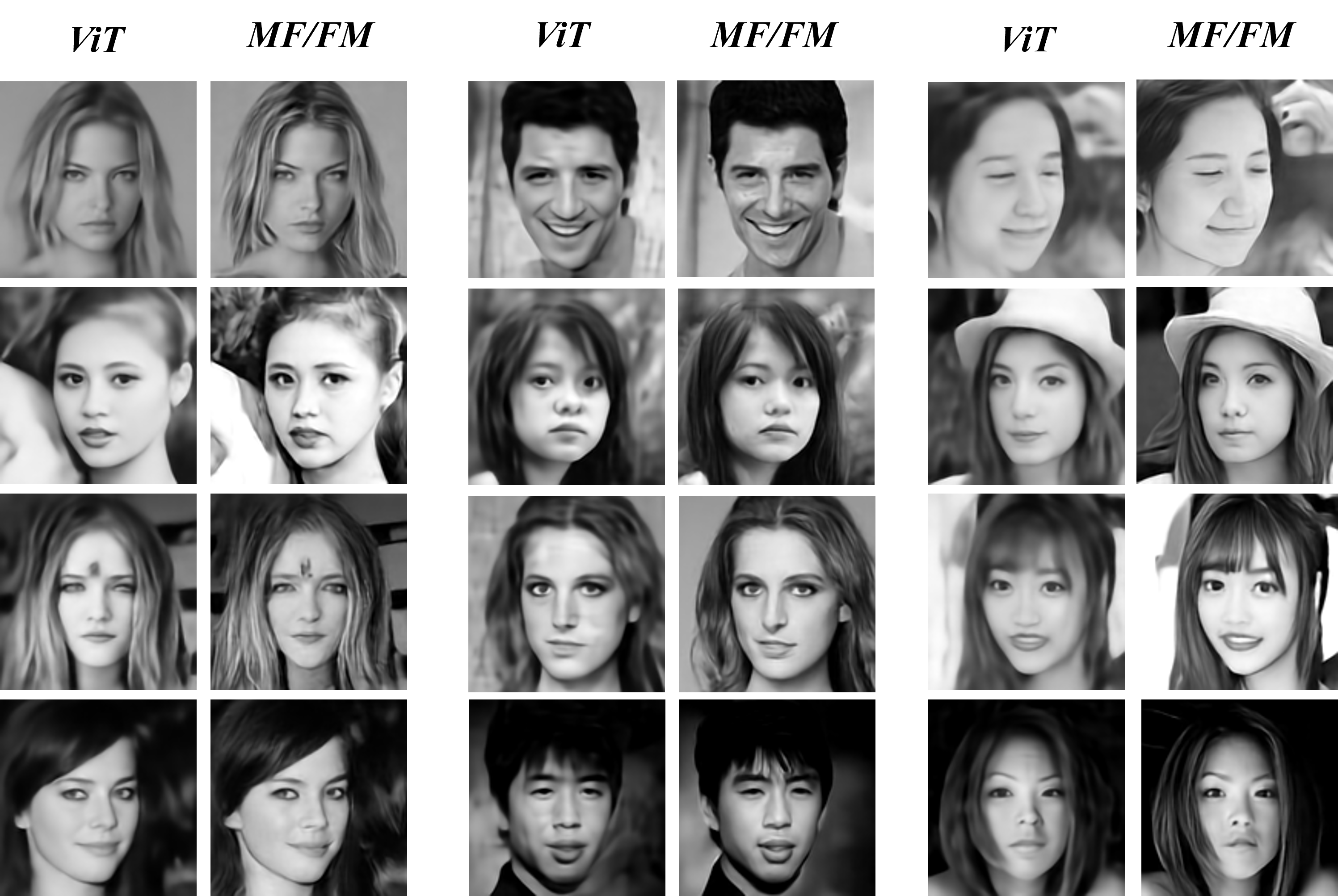} 
  \caption{Regarding the image enhancement performance of MF and FM, the reconstruction time of FM is 20 seconds (NFE=160), while that of MF is 0.238 seconds (NFE=2).} 
  \label{fig:five} 
\end{figure}

In contrast, the probabilistic reconstruction methods we introduce, FM  and MF , can fully leverage the high-quality visual priors embedded in their powerful generative modeling capabilities. These methods can more effectively recover lost high-frequency structural information in images, achieve efficient blind denoising without explicit noise models, and significantly enhance the visual fidelity and perceptual quality of images. Fig. \ref{fig:five} below intuitively demonstrates the comparative results, fully validating the above conclusions.
\end{multicols}

\begin{multicols}{2}
As illustrated in the above figure, the introduction of a probabilistic reconstruction model has effectively enhanced the outcomes of the deterministic reconstruction model. In particular, by leveraging the ControlNet network, we were able to effectively maintain the balance and consistency between the outputs of the two models throughout the generation process. Reconstructed images processed with MF/FM enhancement techniques remain structurally highly consistent with the original ViT reconstruction results, while significantly improving the capability to recover key details: on the one hand, they effectively restore high-frequency details of the image (e.g., textures and edges); on the other hand, they achieve favorable blind denoising performance, collectively elevating both the visual quality and the amount of usable information in the image.  

Both the MF and FM models demonstrate effective image enhancement capabilities, yet they exhibit notable differences in their model characteristics, each with distinct advantages. Specifically, the core strength of FM lies in its simple and straightforward training pipeline, which imposes relatively low demands on training expertise and hardware resources. However, a significant drawback of FM is its slow reconstruction speed: in experiments, a second-order solver (RF-Solver \cite{wangTamingRectifiedFlow2024}) was employed with the number of function evaluations (NFE) set to 160, resulting in a per-image processing time of approximately 20 seconds—far exceeding the threshold for real-time imaging applications. In contrast, the MF model successfully addresses this speed bottleneck, substantially reducing reconstruction time and achieving practical real-time imaging capabilities. Nevertheless, this speed advantage comes at the cost of reduced usability: the training process of the MF model is notably more complex and cumbersome, and it also requires higher computational hardware resources (particularly GPU memory capacity). To accommodate the complex architecture of the MF model and optimize its training efficiency, we configured the DiT model parameters as follows: (depth=24, hidden dim=512, heads=8, patch size=2×2). Additional training configurations are detailed in Table \ref{tab3} below.
\begin{table*}[htbp]
\centering
\caption{Training Configuration}
\label{tab3}
\begin{tabular*}{0.9\textwidth}{@{\extracolsep{\fill}}cccccc@{}}
\toprule
\textbf{Model} & \textbf{Mixed Precision} & \textbf{Flash Attention} & \textbf{Gradient Accumulation} & \textbf{CFG} & \textbf{Epoch} \\
\midrule
FM & Use & Use & No & No & 200 \\
MF & Use & Unusable & Use & No & 700 \\
MF & Use & Unusable & Use & Use & 200 \\
\bottomrule
\end{tabular*}
\end{table*}
As shown in the table, the training process of FM is the most straightforward, requiring no complex training techniques and being compatible with Flash Attention-accelerated \cite{daoFlashattentionFastMemoryefficient2022} training. In contrast, the training of MF is more complex: due to its loss equation involving time derivative calculations, it requires the invocation of the JVP (Jacobian-Vector Product) function, leading to a sharp increase in GPU memory usage; meanwhile, JVP is currently incompatible with Flash Attention acceleration. These factors significantly constrain the usable training batch size. To mitigate this issue, we employed gradient accumulation techniques, achieving a training effect equivalent to a batch size of 200. Additionally, CFG techniques are also critical for the MF model—as they introduce bucket signals as conditional inputs, significantly enhancing both the model's training speed and generation quality. During training, our CFG parameters were set to (w, k) = (2, 0).
\end{multicols}
\begin{figure}[htbp]
  \centering
  \includegraphics[width=0.8\textwidth]{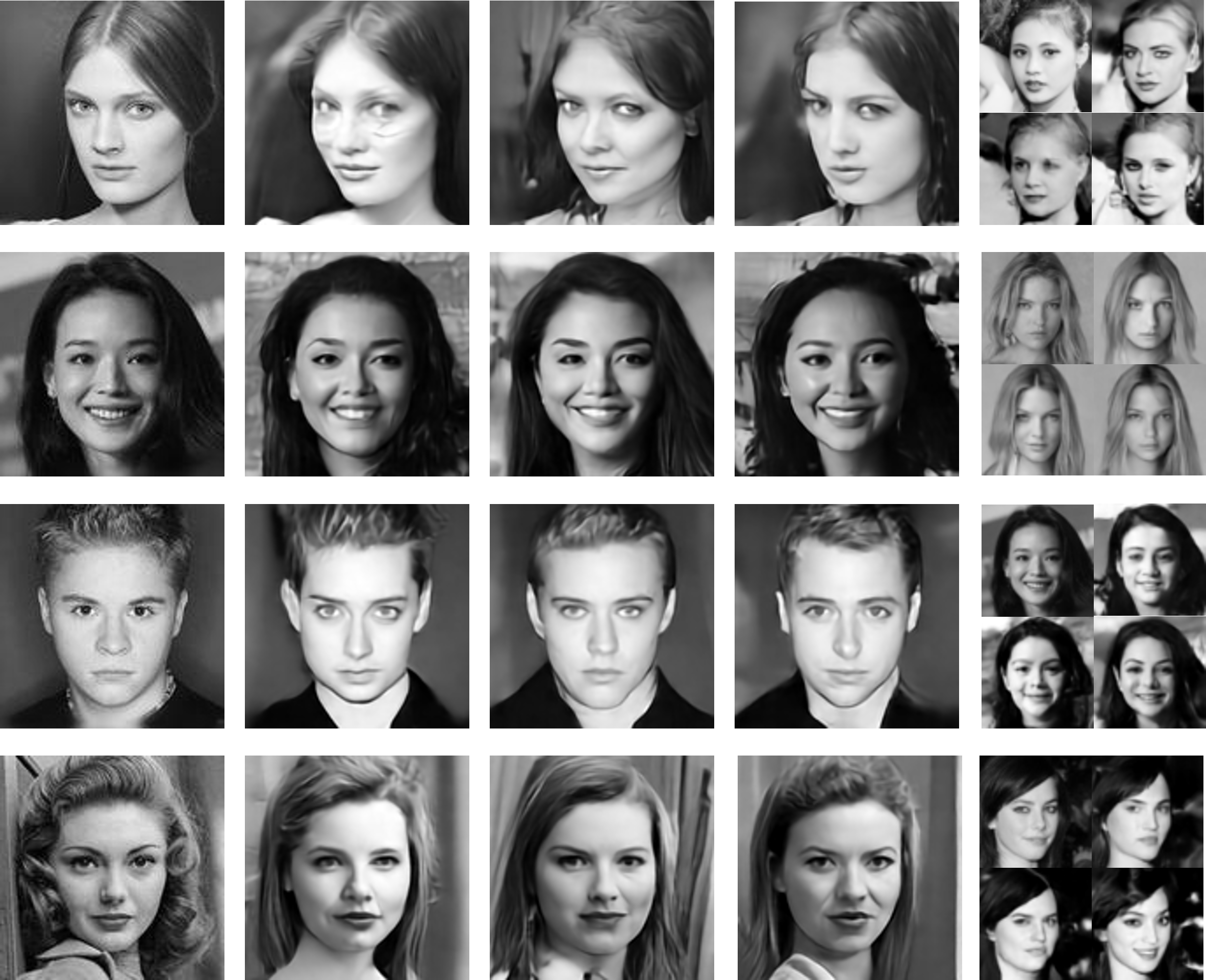} 
  \caption{The diversity generated results utilizing the bucket signal (with the 64×64 image being the CLIP employed for bucket signal encoding).} 
  \label{fig:six} 
\end{figure}
\begin{multicols}{2}
\subsection{Diserity Generation Result}
Previous studies have explored methods for image reconstruction using bucket signals. Building on this foundation, this research further investigates the application of bucket signals in image generation. The findings reveal that bucket signals not only facilitate image reconstruction but also possess natural language-like guidance capabilities, enabling them to direct the generation of images with specific features and thereby expanding their application scope. Relevant experimental results are presented in Fig. \ref{fig:six}. As shown in the figure, when using bucket signals to perform generation tasks, due to the absence of deterministic image guidance, the generated images exhibit a certain degree of diversity. However, for the same set of bucket signals, the generated images consistently share similar features. Additionally, when using bucket signals for generation tasks, the precision requirements for bucket signals are relatively low—even after rounding the bucket signals to the nearest integer (as shown in the bottom row of Fig. \ref{fig:six}), images with similar features can still be generated.

\subsection{Experment Result}
\begin{figure}[H]
  \centering
  \includegraphics[width=0.9\linewidth]{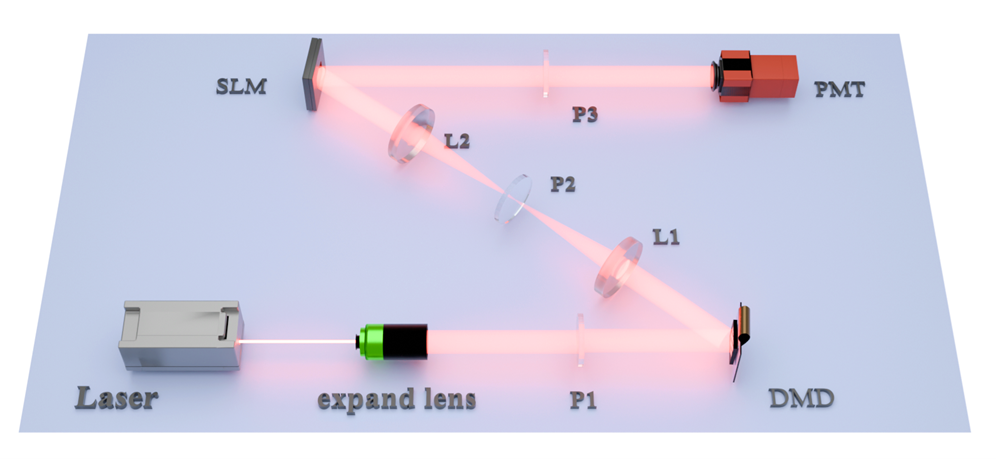}
  \caption{Single-pixel imaging system.}
  \label{fig:seven}
\end{figure}

The optical configuration of the single-pixel imaging system employed in this study is illustrated in Fig. \ref{fig:seven}. The illumination light source of the system utilizes a continuous-wave laser with a wavelength of 532 nm. The laser beam first undergoes beam expansion via a beam expander, then passes through polarizer P1 to form a polarized beam, which is incident on a digital micromirror device (DMD) loaded with a pre-trained speckle pattern to achieve structured light field modulation. The modulated speckle field is transmitted through a 4f optical system composed of lens L1, polarizer P2, and lens L2. This 4f system effectively suppresses environmental noise interference, after which the speckle field carrying modulated information acts on a spatial light modulator (SLM) loaded with the test object. Finally, high-precision bucket detector intensity information is acquired via a photomultiplier tube (PMT).

Subsequently, the Chinese characters ``Dan", ``Xiang", and ``Su" were selected as experimental targets, each with a resolution of 128×128 pixels. The same sampling configuration employed in numerical simulations was adopted for this experiment. The training dataset consisted of an extremely small-scale collection of 6,000 Chinese character images. Both training parameters and hardware setup remained identical to those used in numerical simulations. The imaging results are shown in Fig. \ref{fig:eight} below.
\begin{figure}[H]
  \centering
  \includegraphics[width=0.9\linewidth]{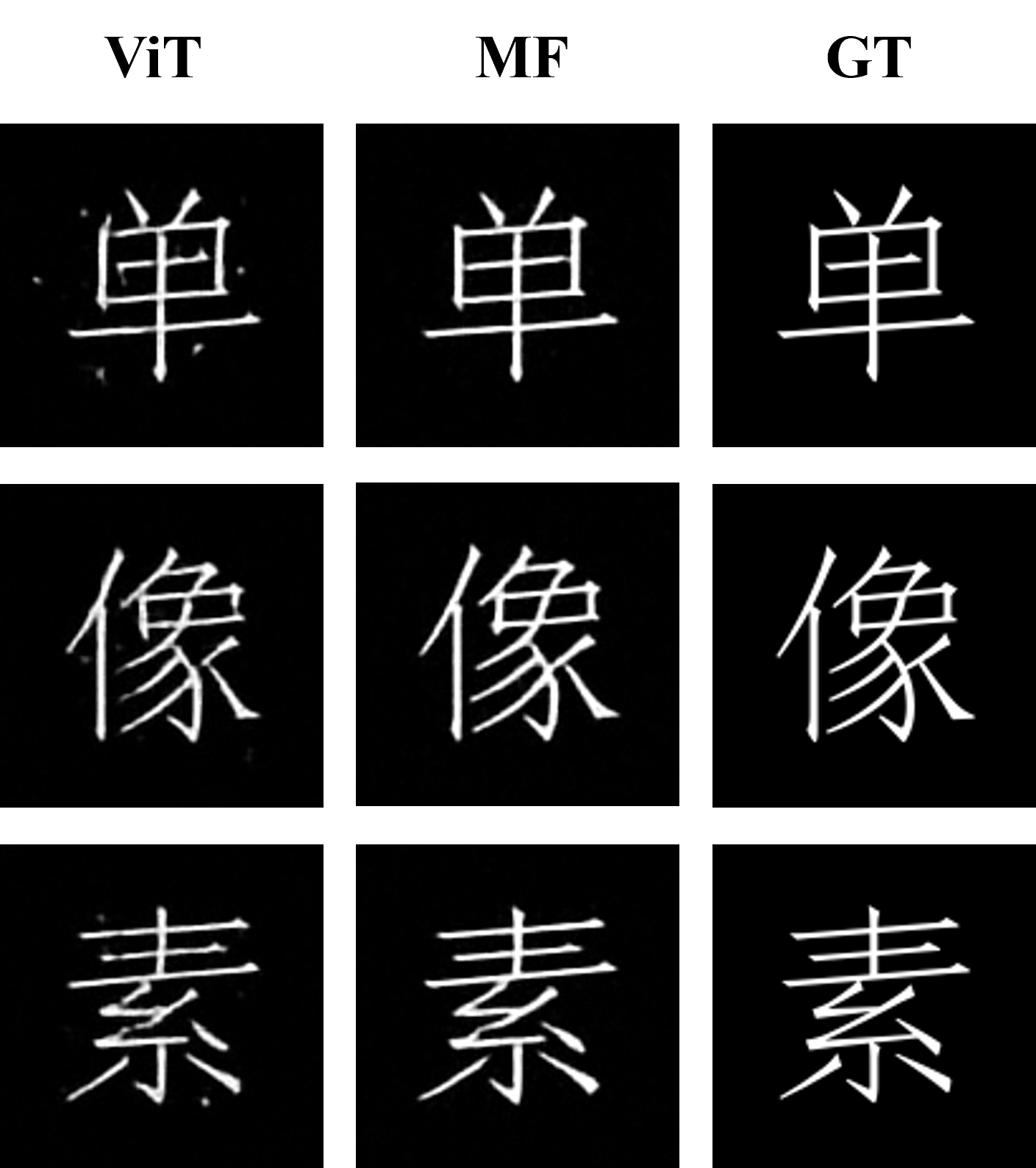}
  \caption{Reconstruction results in experiment.}
  \label{fig:eight}
\end{figure}
Due to environmental noise present in practical experiments that was unaccounted for during model training, certain artifacts are discernible in the ViT reconstruction results. However, by leveraging the powerful generative prior of the MF model, significant denoising and structural restoration were achieved for the ViT based reconstructions. This outcome is particularly challenging for deterministic reconstruction methods.

Subsequent experiments were conducted on the Cartoon Set dataset. A total of 30000 images were selected as the training set for network optimization. Comparative studies were performed against Physics enhanced and DDPMGI approaches, along with diversity generation experiments. The results are shown in Fig. \ref{fig:nine} below.

As shown in Fig. \ref{fig:nine}(a), the physics-enhanced method, being a deterministic reconstruction model, yields inferior image quality compared to generative models. Nevertheless, it preserves the structural features of the original image relatively well. In contrast, DDPMGI, as a generative model, achieves better reconstruction quality but suffers from poor controllability and notable structural deviations from the ground truth. LSSPI, however, combines the strengths of both approaches, striking an effective balance between reconstruction quality and controllability while delivering the best overall visual performance.
\begin{figure}[H]
  \centering
  \includegraphics[width=1\linewidth]{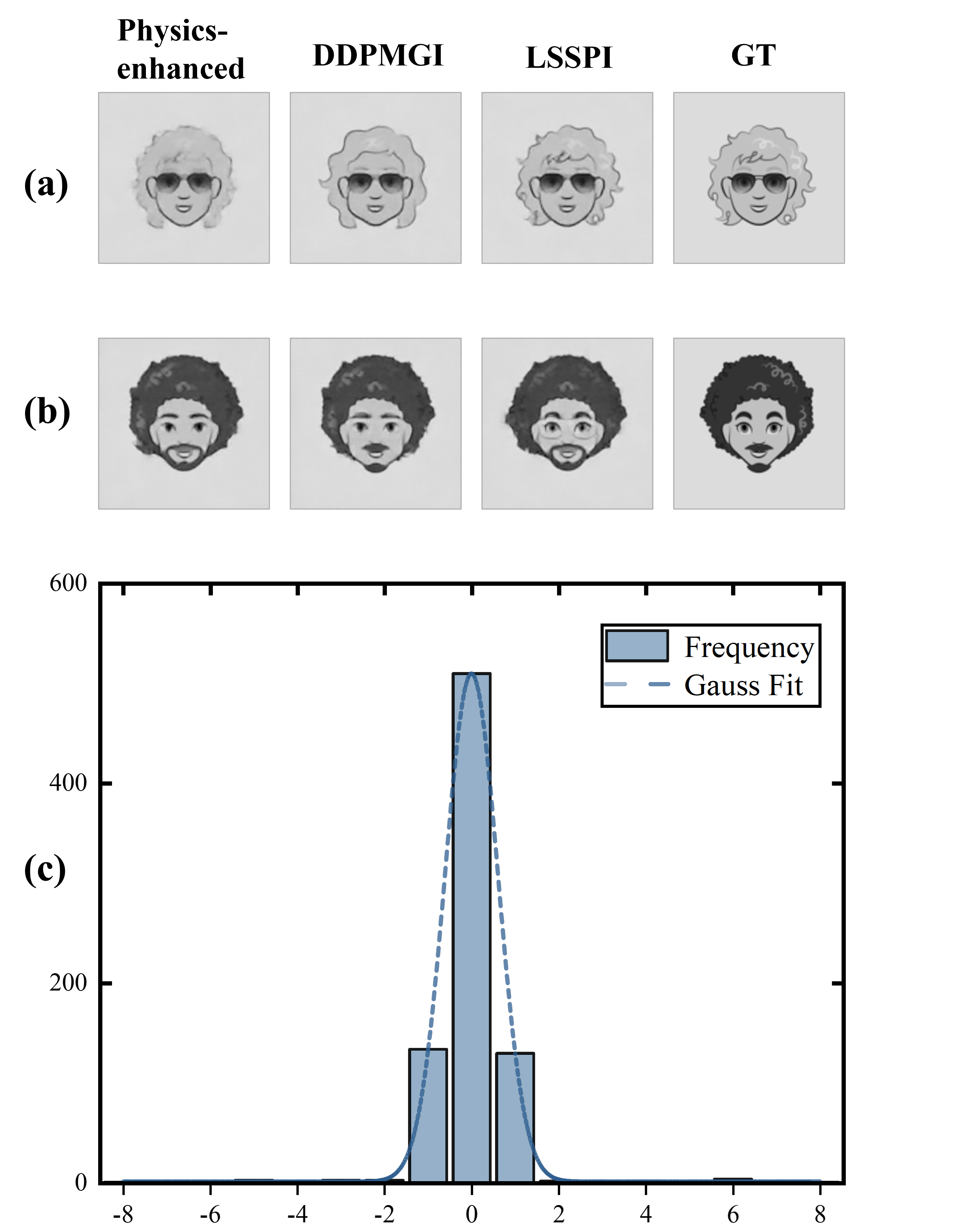}
  \caption{Reconstruction results in experiment. (a) Comparison of reconstruction methods. (b) Generated Results (bucket signals rounded to integer values only). (c) Distribution of bucket signals after rounding}
  \label{fig:nine}
\end{figure}

Fig. \ref{fig:nine}(b) presents the results of the diversity generation experiment, where the bucket signals were rounded to retain only the integer components. The results demonstrate that even under these conditions, the bucket signals remain capable of achieving satisfactory generation performance.
\section{Conclusions}
In this work, we innovatively migrate the single-pixel imaging task to the latent space and propose a novel image reconstruction framework named LSSPI. By combining the strengths of both deterministic and probabilistic reconstruction models within the latent space, LSSPI demonstrates superior performance compared to traditional deep learning networks. Specifically, under equivalent sampling ratios, it achieves reconstructed images with higher signal-to-noise ratio, richer detail preservation, and potential for real-time imaging. Furthermore, LSSPI exhibits blind denoising capability, effectively restoring high-frequency information of images, thereby overcoming the limitations of detail loss and noise interference in low-sampling scenarios typical of conventional methods.

Notably, the latent space migration strategy optimizes both model parameter scale and reconstruction speed, significantly enhancing the practical applicability of LSSPI in low-sampling single-pixel imaging. Despite these advantages, the current results remain limited to small-scale datasets and have not yet been extended to more complex and larger datasets. Additionally, within the LSSPI framework, bucket signals can be utilized not only for image reconstruction but also for image generation. Exploring how to further broaden the application scope of bucket signals presents an important direction for future research.

Moving forward, we will focus on two main aspects: validating the proposed method on diverse and complex datasets including natural landscapes, medical images, and industrial inspection images, and expanding the application range of bucket signals by leveraging more advanced artificial intelligence techniques. These efforts aim to advance single-pixel imaging toward practical implementation.

\bibliographystyle{unsrt} 

\bibliography{main}
\end{multicols}
\end{document}